\documentclass[a4paper]{jpconf}
\usepackage{graphicx}

\newcommand{\beq}{\begin{equation}}
\newcommand{\eeq}{\end{equation}}
\newcommand{\beqn}{\begin{eqnarray}}
\newcommand{\eeqn}{\end{eqnarray}}
\newcommand{\bearr}{\begin{array}}
\newcommand{\enarr}{\end{array}}

\def\bea{\begin{eqnarray}}
\def\eea{\end{eqnarray}}
\def\ba{\begin{array}}
\def\ea{\end{array}}

\begin{document}
\title{ Modules of human micro-RNA co-target network}
\author{Mahashweta Basu$^{1,\dagger}$,Nitai P. Bhattacharyya$^{2}$,P. K. Mohanty$^{1}$}

\address{$^1$Theoretical Condensed Matter Physics Division,\\ Saha Institute of Nuclear Physics,
1/AF Bidhan Nagar, Kolkata 700064, India.}

\address{$^2$Crystallography and Molecular Biology Division,\\ Saha Institute of Nuclear Physics,
1/AF Bidhan Nagar, Kolkata 700064, India.}

\ead{$^{\dagger}$mahashweta.basu@saha.ac.in}

\begin{abstract}
Human micro RNAs (miRNAs) target about $90\%$ of the coding genes and form a complex regulatory network. 
We study the community structure of the  miRNA co-target network  considering 
miRNAs as the nodes which are connected  by weighted links. The weight of link that connects a 
pair of miRNAs denote the total number of  common transcripts  targeted by that pair.  We argue that the 
network consists of about $74$ modules, quite similar to the components (or clusters) obtained earlier 
[Online J Bioinformatics, 10,280 ], indicating that the components of the miRNA co-target 
network are self organized in a way to maximize the modularity.
\end{abstract}


\section{Introduction} 
Micro RNAs  are a class of small single stranded non-coding RNAs, about $20$ to $22$ base long,  which 
interfere with the translation of messenger RNAs (mRNAs) by binding to their $3^\prime$ untranslated regions
 (UTR) \cite{Rajewsky}. Several computational methods 
\cite{comp} have been developed for predicting  the mRNA transcripts which  
are possible targets  of a  particular miRNA. For example,   $711$  nucleotide sequences are  
predicted as miRNAs \cite{miR711}  of human; their  possible targets, ($34525$ in total) 
are listed in the mirBASE database \cite{miRbase}. 
It has been proposed on the basis of theoretical analysis that as large as $90\%$ human genes are
targets of miRNA \cite{Miranda}. Regulation of coding genes by miRNAs in combination
are also experimentally validated \cite{Krek}. 

The abundance of miRNA  and their targets  provide  enormous combinatorial possibilities  for regulation.
Combinatorial regulation of genes by transcription factor (TF) and miRNAs provides higher complex 
programs \cite{Zhou1}.
Recently, taking  TFs  as important mediators of 
miRNA-initiated regulatory effects, it  was shown \cite{Tu} that the  underlying network is 
significantly associated with  multicellular organismal development, 
cell development and cell-cell signaling.  
Combinatorial effect of miRNA modules \cite{Sanghamitra}  has been observed in tumor tissues or cell lines. 
This observation suggests a combinatorial effect of the module associated miRNAs on target gene regulation in selective tumor tissues or cell lines. 
Synergistic network \cite{Xu} of miRNAs reveals that  miRNA modules  associated with diseases are significantly different from  modules of miRNAs that does not involve in disease. 
Possibility of co-regulation of two or more miRNAs in context of gene expressions  and  
relevant biological functions  is, however, least explored.  

Recently Mookherjee {\it et. al.} \cite{mookherjee}  have  analyzed the miRNA co-target network (MCN) of 
\textit{Homo sapiens}, which indicate that  several group of miRNAs (so called {\it clusters})  provide 
most essential regulations.
This topological analysis of miRNA network revealed  that  about $70$ clusters of miRNAs  
co-target the genes, which are involved  in specific pathways. For several clusters, 
all miRNAs belonging to  the cluster  are found to be  maximally expressed in a specific 
tissue.  Further studies \cite{arXiv}, indicate that the clusters are  also disease specific. 
Reorganizing miRNAs into such groups (clusters) helps in identifying cooperative activity of
miRNAs. In fact, from these analysis one can predict   
that, ``if one miRNA from a particular cluster is involved in a  specific biological pathway or cellular function, the other miRNAs belonging to the same cluster are  likely to be involved in the same disease, pathway or function".
 
Detection of  communities, groups, components or clusters have been 
a focus of recent interest  in  context of complex networks. 
Networks like the world wide web \cite{worldweb}, the metabolic network \cite{metabolic}, the  social network
 \cite{social},  protein protein interaction network \cite{ppin} etc. 
do  possess \textit{community structures}, meaning the vertices tend to divide into
groups, with dense connections within the groups and sparse connection existing among the groups. 
These communities act as the functional units of the network; for example `ATP synthesis', `DNA processing', and `cell cycle control' are well known \cite{Zhou}   functional 
modules of  yeast protein-protein interaction network. Evidently, the functional 
properties of an entire network  is quite different from their properties at community level. 

\begin{figure}[h]
 \centering
 \includegraphics[width=7cm]{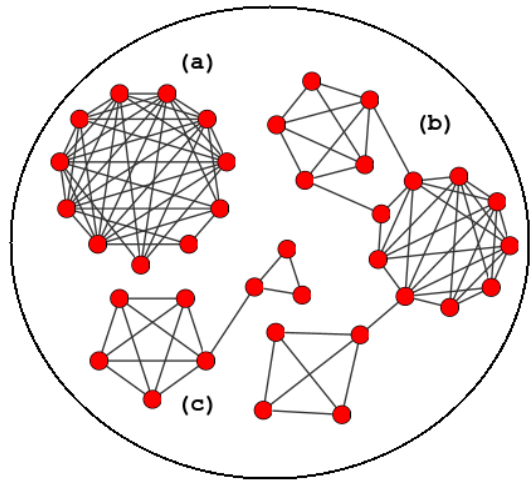}
 \caption{Component versus modules :  Components  (or clusters) are disjoint  sub-graphs of a network (they do not have any common link), whereas modules  may have relatively fewer connections among them. The above network has $3$ components namely (a), (b) and (c) with each component consisting of $1$, $3$ and $2$
 modules respectively. Component (b) has three modules consisting of $4$, $9$ and $5$ vertices, similarly
component (c) consists of two modules comprising of $3$ and $5$ vertices.}
 \label{fig:sch}
\end{figure}

In this article, we study the community structures (modules) of miRNA co-target network of human and 
compare them with the  components (clusters) of miRNAs obtained  earlier \cite{mookherjee}.
Since the components of a network are the only disjoint subgraphs, it is expected that  
the  community structures can be better represented  by  the modules. This is explained 
schematically in Fig. \ref{fig:sch}, where the network has $3$ components and $6$ modules. 

In section \ref{sec:cluster} we briefly review  the relevant features of the 
miRNA co-target network of human and its components (clusters). In 
section \ref{sec:module} we apply the modularization method  introduced by 
Newmann \cite{newman} to analyze this miRNA co-target network and compare the
resulting modules  with the clusters  obtained earlier\cite{mookherjee}. 
 Finally, conclusions are given in section \ref{sec:conclusion}.

\section{Clusters of miRNAs in the co-target network \label{sec:cluster}}

Let us briefly revisit the main ideas and results of  Ref. \cite{mookherjee} to
understand the construction, topology and components of human miRNA co-target network 
consisting of $711$ miRNAs and their $34525$ predicted targets obtained from the
miRBase database (\textit{http://microrna.sanger.ac.uk/, version} $10$). 
For convenience, miRNAs  are given arbitrary identification number $m = 1, 2,\dots i,\dots M$, where  $M=711.$ 
Further, the  miRNA co-target network  was constructed by considering  miRNAs as the nodes and 
joining every pair of miRNAs having one or more common targets with a link. 
The total number of co-targeted transcripts $C_{ij}$ of miRNAs $i$ and $j$  is taken 
to be the weight of the corresponding link. Clearly, the resulting adjacency matrix 
$C$ is symmetric (with diagonal elements $C_{ii}=0$).

Mookherjee et. al. \cite{mookherjee}  have proposed an elegant method for finding the clusters
of miRNAs. Since  substantially large  number of miRNA pairs have  only few co-targets, the 
links  between them have  small weights, and can be erased to obtain a simplified network. 
Let $N_q$ be the number of components  of the  network when all the links having weight less 
than $q$ are erased.  Thus, the network breaks into smaller 
disjoint subgraphs (components) with rate ${dN_q \over dq}$, which is maximum at $q=q^*.$ It was argued  
that among all the subgraphs  of the co-target network obtained at $q^*$,  the largest  one   
is the most important; miRNAs belonging to this subgraph are found to down regulate 
expression of genes involved in several genetic diseases.

To be specific,  the  human  miRNA co-target network breaks into $N_q = 166$ subgraphs at  
$q∗ = 103$, where  the largest  subgraph called $\cal G$ 
contain  $477$ miRNAs.  To determine how miRNAs are organized within 
the subgraph $\cal G$, $q$ is increased further. At $q=160$ the subgraph $\cal G$ breaks up 
into $70$ small clusters (the subgraphs having two or more miRNAs) and $147$ independent miRNAs.   
Out of total $70$, $18$ clusters arise 
from the seed sequence\footnote{The nucleotides $2$ - $7$ of the miRNAs are called seed sequences.}  
similarities and $11$ clusters are organized into the same genomic
region ($5$ inter-genomic clusters  also show  seed sequence similarities). Most of 
the clusters are found to be either pathways, tissues  and/or diseases specific.

 In the following section we aim at investigating the modular structure  of miRNA co-target 
network.  Figure  \ref{fig:sch} schematically describes, why  a network is better represented by its 
modules  than its components (disjoint subgraphs). 

\section{Modular structure of miRNA network\label{sec:module}}

\begin{figure}[h]
 \centering
 \includegraphics[width=8cm]{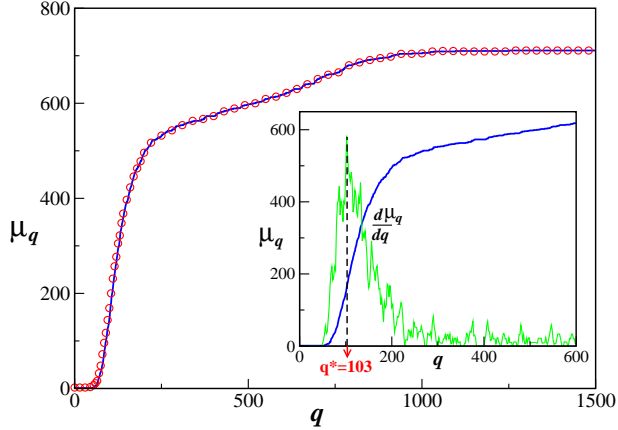}
 \caption{ When all the links having weight less than $q$ are erased, the miRNA co-target network breaks 
into  $N_q$ distinct components comprising of $\mu_q$ modules.  The main figure compares 
$\mu_q$  (solid line) with the $N_q$ (circles). The inset shows that the  ${d \mu_q\over dq}$ is maximum at $q=103$,  
which is same as the value obtained from ${d N_q\over dq}$ earlier  \cite{mookherjee}.}
 \label{fig:Nq}
\end{figure}

The identification of community structure is one of the many challenging problems in various scientific field. 
A large variety of community detection techniques have been developed based on centrality measures, 
link density, percolation theory etc.  Recently, Newman \textit{et. al.}\cite{newman} proposed a method of 
finding community structure of a network based on maximization of  the {\it modularity}. This method 
is further generalized to include weighted networks\cite{weightedNet}.

The most obvious way of finding groups
in a network is to minimize the number of edges connecting the groups.
Simply rearranging the network, such  that only few edges exist between the communities,  
is not enough. Rather one must rearrange it in a way that communities are connected with
fewer than expected edges. One can associate a score called modularity $Q$ \cite{Girvan} 
for each possible partition of a network.
$Q$ is defined up to a multiplicative factor as 
the number of edges present within the groups minus the expected number in an equivalent random
 network. 
Since positive values of $Q$ 
indicates possible presence of community structure, 
one need to look for a partition for which the  modularity is preferably large and positive.

A  good partition of a network can be obtained by  maximizing the  modularity index $Q$ 
defined as follows. Let us consider a  network with  $n$ vertices labeled by $i=1,\dots n$, 
and $m$ links. The corresponding adjacency matrix is $A_{ij}$. 
Let the degree of each vertex $i$ is $k_i= \sum_j A_{ij}$, thus $m={1\over2} \sum_i k_i$. 
If the  network is to be partitioned into two groups, one  associates a quantity $s_i$ which 
takes  a value $s_i=+1 (-1) $  if vertex $i$   belongs to group $1$ (group $2$). 

Correspondingly the  modularity is given by  
\begin{equation}
Q={1\over 4m}\sum_{ij}\left(A_{ij}- {k_i k_j\over 2m}  \right)(1+s_i s_j),
\label{eq:mdl}
\end{equation}
where $k_ik_j/2m$ is the expected number of links  between $i$ 
and $j$, if edges were placed at random. The term $(1+s_i s_j)$  is $0$  ($1$)  if vertices 
$i$ and $j$ belong to different (same) group; this assures that $Q$ is maximum when two 
groups are  connected  by  smaller than expected number of links. In the following we apply this procedure to obtain  modular structure of MCN. 

MCN is a undirected weighted network, where the weight of the link $C_{ij}$ corresponds to  the number
 of transcripts  being co-targeted by the concerned pair of miRNAs $i$ and $j$. The diagonal elements  
are taken $C_{ii}=0$, as usual.  
It has been pointed out  in Ref. \cite{mookherjee}  that the weights  vary widely between  
$1$ to $1253$, indicating that  most of the links with small weights can be erased 
to obtain a simpler network. However,  the connectivity of the 
network changes when links having weight less than a predefined value $q$  
are erased. Taking the adjacency matrix $C^q$, defines as
 \begin{equation}
C_{ij}^q = \left\{ 
\begin{array}{cc} 
1 & \textrm{if~} C_{ij}>q \cr 0&  \rm{ otherwise,} \end{array}\right. 
\nonumber
\end{equation}
Mookherjee  \textit{et. al.} \cite{mookherjee} have calculated the  number of 
components $N_q$ by varying $q$. Since,  this adjacency matrix $C^q$ is unweighted
(as it keeps the  information of connectivity ignoring the actual weights)  
one can apply the idea of modularity maximization\cite{newman} to detect the communities 
or modules present there. Let $\mu_q$ be the  total number of modules of $C^q$.
Clearly $ \mu_q \ge N_q$, as  each component can either have one module, \textit{i.e.} itself, 
or it can break into two or more modules.  
In Fig. \ref{fig:Nq}  we have compared $\mu_q$, obtained from modularization 
methods,  with the components $N_q$ obtained earlier.  It is evident that  
$\mu_q \simeq N_q$; a negligible small  positive  difference is  not visible 
in the figure.  This  brings us to conclude that  the components of the network 
are self organized  in a way that modularity  (given by Eq. (\ref{eq:mdl})) is 
maximized.

   \begin{figure}[t]
 \centering
 \includegraphics[width=10cm]{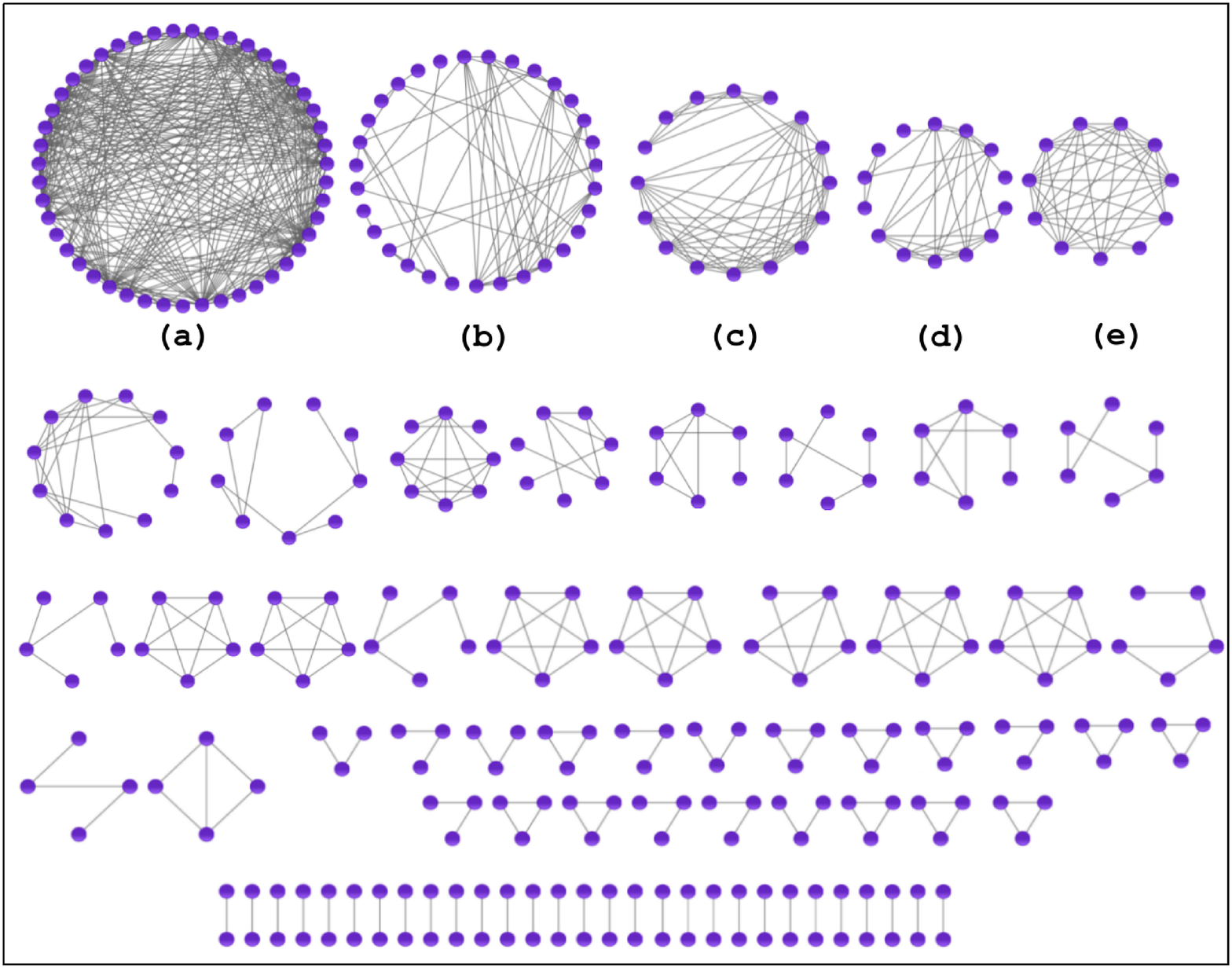}
 \caption{Components of $\cal G$ (which has $477$ miRNAs) :  Only the clusters (components of 
size larger than $1$) are shown.  Each of the large clusters with size $(7,8,9,14,16,31,47)$ appear
once. The other $63$ clusters comprises of $(2;29)$, $(3;21)$, $(4;2)$, $(5;7)$, $(6;2)$,$(11;2)$, 
where  $(m;n)$  denotes  that cluster of size $m$ appears $n$ times.}
 \label{fig:clus477}
\end{figure}


We also find that ${d \mu_q\over dq}$ is 
maximum at $q=103$,  which is the value obtained from ${d N_q\over dq}$ earlier \cite{mookherjee}. 
Now let us have a closer look  at  the size of the components  and that of the modules 
obtained at $q*= 103$. This is listed in Table-I. Note, that  the modularity maximization algorithm  
organizes the network into several small modules and few  moderate size modules 
as $26, 37, 57, 98, 101$. Whereas in terms of components the network breaks into few clusters 
of small sizes ({\it e.g.} $2, 3, 4, 5, 7$) along with a {\it giant} cluster ($\cal G$) 
of size $477$ \cite{mookherjee}. 
\begin{table}
\label{table:I}
\centering
 { \tiny
\begin{tabular}{|c|c|c|c|c|c|c|c|c|c|c|c|c|c|c|c|c|c|c|c|}
\hline
{\bf Size} & 2 & 3 & 4 & 5 & 6 & 7 & 8 & 9 & 11 & 12 & 13 & 14 & 17 & 26 & 37 & 57 & 98 & 101 & 477 \\ 
\hline 
{\bf No. of clusters}  & 24 & 7 & 1 & 2 & 0 & 3 & 0 & 0 & 0 & 0 & 0 & 0 & 0 & 0 & 0 & 0 & 0 & 0 & 1 \\
\hline
{\bf No. of modules} & 26 & 8 & 2 & 3 & 1 & 4 & 1 & 2 & 2 & 2 & 2 & 1 & 1 & 1 & 1 & 1 & 1 & 1 & 0 \\
\hline
\end{tabular}
}
 \caption{Comparison of  number of modules  of the miRNA network  with the number of components,  at $q=q*=103$.}
\end{table}
Evidently  at $q*=103$, MCN has {\it one} distinctly large component 
containing $477$ miRNAs, compared to the moderate size modules those appear 
with competitive sizes ($101$ and $98$). 
The largest component must have been broken  into these smaller modules. 
Thus, as far as `identifying a large set of relevant miRNAs' (one like ${\cal G}$) is concerned, 
one can reliably consider the component ${\cal G}$ as
\begin{figure}[t]
 \centering
 \includegraphics[width=10cm]{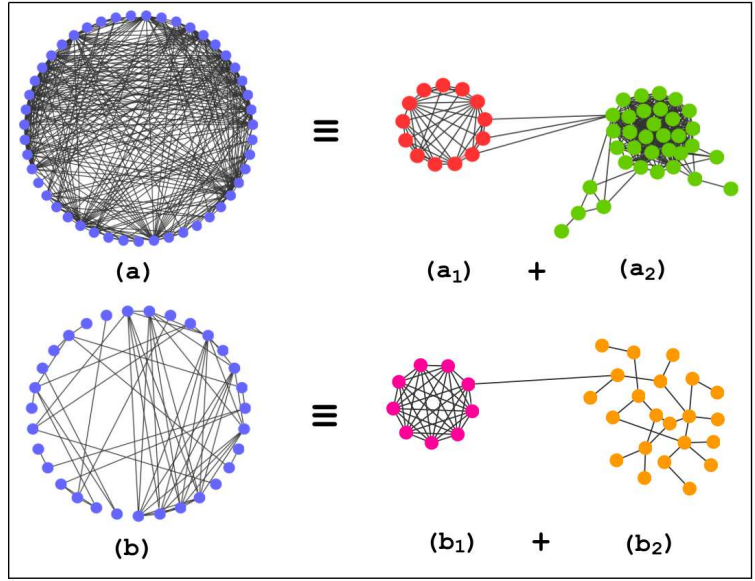} 
 \caption{ The cluster (a)  and (b) of Fig. \ref{fig:clus477} are redrawn to visualize the 
the community structure obtained through modularization. Clearly (a)  which contains 
$47$ miRNAs has two modules $a_1$ ($34$ miRNAs) and $a_2$ ($13$ miRNAs), and  (b) has two modules   
$b_1$ ($22$ miRNAs) and $b_2$ ($9$ miRNAs).}  
 \label{fig:clus47}
\end{figure}
the optimal set of miRNAs, which 
co-regulate the gene expressions. Further,  to understand  how miRNAs are 
organized within $\cal G$, we calculate 
its modules  by taking $q=160$,  which is the same value of $q$  used in \cite{mookherjee},   
to obtains the  clusters  (in total  $70$). All the  components of ${\cal G}$ 
having two or more miRNAs (referred to as clusters), are shown in 
Fig. \ref{fig:clus477} in decreasing order of their sizes. The  first 
five,  named as (a) to (e) have  $47,31,16,14$ and $9$ miRNAs respectively.

It would be interesting to look  at the community structure of ${\cal G}$ at this value of 
$q=160$. Using the modularization algorithm \cite{newman}, we find that ${\cal G}$ contains 
$72$ modules (total $330$ miRNAs) and  $147$ single miRNAs. Whereas in terms of components, 
${\cal G}$  had  $70$ clusters (total $330$ miRNAs) and $147$ independent miRNAs \cite{mookherjee}.
The  detailed study of modules reveals that only two of the  $70$ clusters are broken into smaller 
modules : cluster (a) in Fig. \ref{fig:clus477} with $47$ miRNA, has two modules 
$a_1$ and $a_2$ of size $34$ and $13$ respectively, cluster (b) with $31$ miRNAs, 
breaks into two modules  $b_1$ ($22$ miRNAs) $b_2$ ($9$ miRNAs).  Such modular structures 
of (a) and (b) were not apparent in Fig. \ref{fig:clus477}; we redraw these  graphs  keeping 
all miRNAs in same module close to each other. The resulting graph (Fig. \ref{fig:clus47} (a) and (b)) 
clearly show the existence of modular structures. 

In summary, the community structure in these networks are very similar to the 
components  (or clusters)  obtained earlier in \cite{mookherjee}. Only  few large components  show 
  further small sub-structures, 
indicating that the  existing components of MCN  are already optimally modularized.  
Implications of these results will be discussed in  section \ref{sec:conclusion}.

Few comments are in order.   It is quite evident that  cluster (c) 
in Fig. \ref{fig:clus477}, containing $16$ miRNAs,   might have sub-structures of 
size $11$ and $5$, which could not be obtained when the  modularization algorithm is 
applied to  the  un-weighted graph ${\cal G}$  containing  $477$ miRNAs. In this 
analysis, the actual weights were ignored, $i.e.,$ all links having weight more 
than $160$ are considered  identical irrespective of  their actual weight. 
When we keep these weights and  use the modified version of the algorithm \cite{weightedNet}, 
that works for  weighted networks, the  cluster (c) shows the predicted substructures. 
In addition, some other modules, such as $a_2$  which has $34$ miRNAs  also show further 
sub-structures  of size $30$ and $4$.  These four nodes, turns out to be those 
shown in  the left side of  $a_2$ in Fig. \ref{fig:clus47}.

It appears that Newman's algorithm, both for un-weighted and  weighted network, 
provides only the sub-structures  of large components. This is because, modularity  of the network 
is inversely proportional to  the total number of links ( see Eq. (\ref{eq:mdl})). Thus, 
the total modularity of a network with many components is not substantially 
altered by re-structuring the small components  into smaller sub-structures. It is only, 
the re-structuring of larger components which can change the modularity appreciably.  To overcome 
this difficulty, one must find modular structures of individual components, instead of 
looking at the community structure of the whole network.  

\section{Conclusion\label{sec:conclusion}} 
To our surprise, the community structure of human miRNA co-target  networks is very 
similar to the existing components  or clusters. Only  few large components  show   smaller   sub-structures. 
Most of the components  do not show any further substructures, indicating that the miRNA co-target 
network inherently consists of optimally  modularized structures. It is quite  possible that, 
 during the evolution of miRNAs,  first the the modular structures are formed, optimized and 
then they join with other modules to provide  essential regulation for  complex life structures. 
Further study in these directions is required  to verify such hypothesis.

\end{document}